\magnification=1200
\nopagenumbers
\input boxedeps.tex 
\input boxedeps.cfg 
\SetDefaultEPSFScale{600}

\centerline {\bf {Three Dimensional Gravity in the Presence of Scalar Fields}}
\vskip .7truein
\centerline {Tolga Birkandan, M.Horta\c csu * }
\vskip .4truein
\centerline {  Physics Department, Faculty of Sciences and Letters}
\vskip .1truein
\centerline { ITU 80626 Maslak, Istanbul, Turkey }
\vskip 1.3truein
\centerline {\bf { Abstract}}
We study a scalar field in curved space in three dimensions.  We obtain a static perturbative solution
 and show that this solution satisfies the exact equations in the asymptotic region at infinity. 
The new solution gives rise to a singularity in the curvature scalar at the  origin.  Our solution, however,  necessitates the excising the region near the origin, thus this naked singularity is avoided.

\vskip .3truein
\noindent
key words: black holes, Choptuik formation,  
\vskip .5truein
\noindent
* hortacsu@itu.edu.tr
\baselineskip=18pt
\footline={\centerline {\folio}}
\pageno=1
\vfill\eject
\noindent {\bf{ INTRODUCTION}}
\bigskip
\noindent

After Choptuik's interesting discovery $^{/1 }$ of the mechanism of the
emergence of black holes , many papers were written $^{/2  }$
on the transition of curved space into blackholes through scalar particles
which are injected towards them.
In this phenomena there is a parameter, which acts like the order parameter
in phase transitions, which decides whether a black hole is
going to be formed or not .  Often this parameter is the amplitude of the wave
which is injected from infinity to the enter of the coordinate system.
If we start using  a scalar field with a  minute amplitude as the initial form
of the particle injected, no singularity is formed in the metric
pointing to the existence of a black hole .  As the amplitude of the scalar
field is increased, we  find a critical value , when exceeded , giving rise
to the formation of a black hole.
This phenomena was later investigated using the BTZ black hole $^{/3}$ as an example.
Numerical and analytical work in this direction are given in references 4-9.

In this note we study the BTZ system along the lines of Choptuik et al's work $^{/4}$
 .  We first study the static case with no scalar field. 
We, then, add the scalar field perturbatively around the zero solution, perform a
perturbation expansion and obtain the solution both for the metric and for the scalar
field by this method.  We see that there is
a singularity only at one end of our domain in the perturbative approach.
We , then, change to new coordinates where we interprete the singularity as
one occurring at the origin. Since our surface is trapped before reaching this point, the "naked 
singularity " is avoided.

We think our results are a further evidence along the line given by the
Choptuik and the subsequent work.  We also point to the emergence of
"hypergeometric functions" of some sort in our solutions, similar to the case
as in Birmingham's work $^{/10  }$ , where  solutions of the Klein-Gordon equation
in the background  metric of the BTZ black hole  were studied. We know that
the presence of hypergeometric equations is a sign of the presence of conformal
symmetry in the problem, in an open or disguised fashion, as they appear in many cases
since the original solution of 't Hooft $^{/11 }$ 
, for the fluctuations in the background
of Yang-Mills instantons . Another well-known example is the seminal paper by the Russian group $^{/12}$
which was on 2-d conformal symmetry, where the solutions were given in terms of hypergeometric
functions.  The same functions  arise $^{/13}$  for N=2 SUSY $SU(2)$ Yang-Mills, in  Seiberg-Witten $^{/14}$ work.  The conformal structure of the BTZ 
solution is also seen in Birmingham's work $^{/10}$ as well as the work of Henneaux et al $^{/ 2}$ where this solution is shown
to be related to the two-dimensional system with right and left  Virasoro structure.
\vskip .1truein
{\bf{CHOPTUIK-PRETORIUS COORDINATES}}
\vskip .1truein
We start with the equations given by Pretorius and Choptuik $^{/4}$
for a metric given by 
$$ ds^2= {{e^{2A(r)}}\over {cos^2 (r) }} (dr^2 - dt^2) + tan^2 (r) e^{2B(r)} d\theta^2 .\eqno {1}$$
Here the full space is mapped into the interval $ 0<r< \pi/2$. 
We choose the cosmological constant $\Lambda $ equal to -1 and scale $r$ so that it is a dimensionless parameter.  We take the 
static and the spherically symmetric  case, where 
$ A, B $ are functions of r only. Since we are treating the non-rotating case,
$\theta$ independence is justified. 

If a scalar, static particle is coupled to the metric, the Einstein equations, 
with the cosmological constant read $^{/4}$
$$ A_{,rr} + {{(1-e^{2A})} \over { cos^2 (r)}} +  2\pi \phi_{,r}^2 =0 , \eqno {2} $$
$$B _{,rr}+ B_{,r} \left( B_{,r} +{4 \over {sin(2r)}}\right) + {2{(1-e^{2A})} \over {cos^2 (r)}} 
=0 , \eqno {3} $$
$$ B_{,rr}+B_{,r} \left( B_{,r}-A_{,r}+2{{(1+ cos^2 (r))} \over {sin (2r) }}\right)-
{2A_{,r} \over { sin (2r)}} + {{(1-e^{2A}) } \over {cos^2(r)}} + 2\pi \phi_{,r}^2 =0,\eqno {4} $$
$$ [ tan r e^B \phi_{,r} ]_{,r} =0 .\eqno {5} $$

This system of equations have the set of solutions 
$$ \phi=0, A= -log (sin r), B= log({{cos 2r} \over {2sin^2 r}}) .\eqno {6}$$
The expressions for $A$ and $\phi$ were given in reference 4   .
We see that our original domain, $ 0< r < \pi/2 $ is halved.  Due to the singularity in $B$,
 we can use only the region where $0< r < \pi/4 $.

We perform a simple perturbation expansion at this point treating the above set of solutions as the zeroth order term.
$$ A= A_0+\epsilon A_1 + \epsilon^2 A_2 + ......   ,\eqno{7}$$
$$ B= B_0+\epsilon B_1 + \epsilon^2 B_2 + .......  , \eqno{8}$$
$$ \phi = \epsilon \phi_1+ \epsilon^2 \phi_2 +..... \eqno {9}$$
Equation ( 5 ) gives us $ \phi_{1,r} = C tan 2r $ where C is a constant of integration.  We choose $C=1/2$.
The equation (  2 )reduces at this order to 
$$ A_{,rr} - {8A \over {sin^2 (2r)}} =0 \eqno{10}$$
which can be reduced to an equation of the hypergeometric type.  A simple calculation shows that 
we have a special form of the  hypergeometric equation, yielding 
$$ sin^{-1} (2r) _2F_1(-1/2,-1/2|-1/2|sin^2 (2r) )\eqno {11}$$ for one solution, and $$ sin^2 (2r)
_2F_1(1,1 | 5/2| sin^2 (2r)\eqno {12} $$ for the other.

These special forms of the hypergeometric function reduce to 
$$ A^1_1= cot (2r) ,\eqno{13}$$
$$ A^2_1 = 3( 1-2r cot 2r) \eqno {14}  
$$
respectively.  For the former solution $ A^1_1$ we get two solutions 
$$ B_1^1 =  cot 2r  + tan 2r,\eqno {15} $$
$$ B_1^2 =  -6r B_1^1 . \eqno {16}$$

By extending the perturbative analysis, for $A$ we get a differential equation,  again of hypergeometric
type , with an inhomogenous term, for $A$  which can be easily integrated.  
We are rather interested in the solution for $\phi$, to study its singularity structure.  To second order in $\epsilon$
$$ \phi_{2,r} = {{12r-3 sin (4r)}\over { 2 cos^2 (2r)}}. \eqno {17} $$
Here we used the solution given by the set $ A_1^2, B_1^2$, anticipating the result we find in the next section, where it is shown that the solution $ A_1^1, B_1^2$ has undesirable  properties and wrong asymptotics, whereas the former set is the "physical" solution.
\vskip .1truein
\noindent
{\bf{ NEW COORDINATES}}
\vskip .1truein

Our results may be interpreted better if we transform our original coordinates to 
$$ \overline {r} = tan (r) e^{B(r)} .\eqno {18}$$
Then our metric is transformed into 
$$ ds^2 = -(-M+ \overline {r}^2 )d \overline {t}^2 + {{d\overline {r}^2}\over {(-M+\overline {r}^2)}} +\overline{r}^2d\theta^2, \eqno {19}$$
where 
$$ -M= e^{2(B-A)}[ sec^2(r) + 2 tan (r) B_{,r} + sin^2(r) B_{,r}^2 ] - tan^2 (r) e^{2B} \eqno{20} $$ 
as given in reference (4).  Using our zeroth order solution for $A$, we find  
$\overline {r} =  cot (2r) + O(\epsilon) .$  This transformation maps our original domain, $ 0 < r < \pi/4 $, into 
$ \infty >  \overline {r} > 0 $.  For the zeroth order solution 
$ -M= 1 $, a constant. We see that our solution corresponds to the AdS solution, known for this system.  If we use the solution $B= log({{cos 2r} \over {2\alpha sin^2 r}})$, where $0<\alpha<1$, we get the solutions corresponding to conical singularities $^{/15}$.   The BTZ solution $^{/3}$ is obtained if we set $\alpha=-i$ which necessitates  a reparametrization using hyperbolic functions . In the parametrization of the metric, eq. (1), hyperbolic cosine and hyperbolic tangent functions replace trigonometric cosine and trigonometric tangent respectively.  The presence of $i$ in the expression for $B$ introduces the nesessary minus sign for $-M$, eq. (22), while replacing $tan (r)$ by $tanh(r)$ in the metric, eq. (1), retains the original signature .   

We, then, use our perturbative solutions in the presence of the scalar field. At first order in $\epsilon$, we find we have two solutions.  If we take the solution set  $A^1_1, B^1_1$, we find
$$ -M=1+32\epsilon[ {{1}\over{\overline {r}}}- {\overline {r}}^3].\eqno {21}$$
This expression diverges both at $\overline {r}=0$and $\overline {r}=\pi/4$ in an undesired fashion, so is discarded.
For the latter solution set, $A^2_1, B^2_1$,
$$ -M = 1 - 96\epsilon [ (1+\overline {r}^2)+ cot^{-1}
(\overline {r})(-\overline {r}^3+{{1}\over{\overline {r}}})],\eqno {22} $$
an expression which diverges at the origin, but this time with the correct sign.  At $\overline {r}=0 (r=\pi/4)$, $-M$ diverges to minus infinity.  We excise the domain when $-M+\overline {r}^2$ equals zero.

When the scalar field is calculated in terms of the new variable, we find 
$$\phi= 1/2 log \left({{(1+{\overline  {r}}^2)^{1/2}}\over {\overline  {r}}}\right). \eqno {23} $$
For $\overline {r} = 0 $, the scalar field $\phi$ is proportional to $log (\overline {r} ) $, as pointed out by Garfinkle $^{/8}$ and Burko $^{/9}$.
When $\overline {r} $ goes to infinity, $r=0$, $\phi$ goes to zero.  
The similar behaviour persists at second order in $\epsilon$, where 
$$  \phi_{2,r} = -3{{1}\over { \overline {r}}}- cot^{-1} (\overline {r}) ({{\overline {r}^2 +1} \over {\overline {r}^2}}).\eqno {24}$$
We see that the divergence is severer when $\overline {r}$ goes to zero.

The fact that $-M$ goes to minus values as $\overline {r}$ approaches zero,  signals the presence of a black hole around the  origin. We can not tolerate $-M+\overline {r}^2 $ being null. We excise the space at the  value of $\overline {r} $ where $-M+\overline {r}^2 $ equals zero. There is only one root of the equation 
$$ -M+{\overline {r}}^2=0, \eqno {25}$$ giving the approximate condition 
$$ \overline {r} > 24\pi \epsilon \left(1-(24)^2 {\pi}^2 {\epsilon}^2)\right)+O({\epsilon}^4) .\eqno {26}$$ 
This will also prevent the curvature singularity which will occur at $\overline {r}=0$.  We have the scalar curvature made out of two parts , the finite part corresponding to the AdS solution and a singular part coming from the perturbative solution.
$$ R = -6+{{\epsilon \pi}\over {\overline {r}^2}} [{{1}\over {1+\overline {r}^2}}],\eqno {27} $$
where $R$ is the curvature scalar $^{/4}$.

To detect whether black-hole is formed or not, we use a second test given in reference (4), by checking the condition for "trapped surfaces". In this reference, " trapped surface" is defined " to be surfaces where the expansion of the outgoing null curves normal to the surface is negative ".  The condition for this to happen is given at the same place as
$$ S= 1+ sin (r)cos (r) B_{,r} < 0 , \eqno {28}$$ 
as applied to our case.
We study whether this constraint is satisfied for our solutions.
To zeroth order in $\epsilon$ this condition reads 
$$ 2S = -{{(1+(\overline{r}^2)^{1/2}}\over{\overline {r}}}. \eqno {29}$$
We see that "the surface is trapped" for $\overline{r} = 0 $ ($r=\pi/4$). 
For $r=0$ i.e. $\overline {r}$ approaching infinity, we get $S = - {{4}\over { \overline {r}^2}}$ which goes to zero from below.  If we use the first order solution, however, the correction  changes the situation in the first case, whereas it does not change the result in the second case.. For the first solution, we have
$$ S= 1+ sin(r) cos (r) B_{,r} =$$
$$ -[(1+(\overline{r})^2)^{1/2}\left({{1}\over {\overline{r}}}+\epsilon (1-{{1}\over {(\overline{r})^2}})\right) ] .\eqno {30}$$
For $\overline{r}/4 < \epsilon $ there is no "trapped surface.  We can come all the way to the curvature singularity.  This solution is discarded also for the above reason.
If we use the second solution, however, we get 
$$ S=
-(1+(\overline{r})^2)^{1/2}({{1}\over {\overline{r}}})$$
$$-3\epsilon\left[(1+(\overline{r})^2)^{-1/2}\left(\left(+\overline{r}+{{1}\over{\overline{r}}}\right)+cot^{-1}(\overline{r})\left({-\overline{r}}^2+{{1}\over{{\overline{r}}^2}}\right)\right)\right].\eqno {31}$$
As $\overline{r}$ goes to zero, the surface is trapped.  The particle is not allowed to come close to the coordinate independent curvature singularity at the origin. 
\vskip .1truein

\noindent
{\bf{ASYMPTOTIC ANALYSIS AND DISCUSSION}}
\vskip .1truein
Here we use our approximate solutions given above, eq.s (15,16,23,24) and check whether they satisfy the set of exact equations, eq.s(2-5).The figures 1 to 4 show that although our approximate solutions do not satisfy the equations as $ \overline{r}$ goes to zero, the behaviour as $\overline{r}$ goes to infinity of these figures clearly show that asymptotically these solutions tend to be exact.   We thus see that we can obtain approximate solutions to the equations of motion and the constraint equations,  which approach exact solutions in the asymptotic region.  Our tests, mainly the fact that $-M+\overline{r}^2$ goes through zero tells us that a blackhole is formed as $\overline{r}$ goes to the origin (figure 5).  We excise the space at the  point where $ -M+\overline{r}^2$ equals to zero.  At this point note that our first set of approximate solutions, eq.s ( 13,15) do not give the correct asymptotic behaviour, so are discarded.

From these figures we see that our approximate solution is no longer reliable as $\overline{r}$ goes to zero. Since we excise our space in this region and the solution has the correct asymptotics as $\overline{r}$ goes to infinity, we think that the message our approximate solution conveys, i.e. the presence of a blackhole at the origin, is correct.

     Further work along these lines can be done using the rotating solution of the Teitelboim and collaborators  $^{/16}$.
\vskip .2truein

\noindent 
{\bf{ Acknowledgement: }}
We thank very illuminating discussions with Professors Alikram Aliev and Ne\c se \" Ozdemir.  This work is partially supported by TUBITAK, the Scientific and Technical Research Council of Turkey.  M.H. is also supported by TUBA, The Academy of Sciences of Turkey.

\vskip 1 truein
\noindent
{{\bf{REFERENCES}}}

\item {1.}   Matthew W. Choptuik, Physical Review Letters,{\bf {40}} (1993) 9;

\item {2.}   For a review see: D. Birmingham, Ivo Sachs and Siddartha Sen, hep-th/0102155; a most recent work is by Marc Henneaux, Cristian Martinez, Ricardo Troncoso and Jorge Zanelli, hep-th/0202270;

\item {3.} M.Banados, C.Teitelboim and Jorge Zanelli, Physical Review Letters, {\bf{69}} (1992) 1849; also, M.Banados, M.Henneaux, C.Teitelboim and J.Zanelli, Physical Review, {\bf{D48}} (1993) 1506.

\item {4.}   Frans Pretorius and Matthew W. Choptuik, Physical Review {\bf{D 62}} (2000) 124012;

\item {5.}   Viqar Husain and Michel Olivier, Classical and Quantum Gravity,{\bf{ 18}} (2001) L1;

\item {6.}   Andrei V. Frolov, Classical and Quantum Gravity, {\bf{16}} (1999) 407;

\item {7.}   G\' erard Cl\' ement and Alessandro Fabbri, gr-qc/0101073;

\item {8.}   D. Garfinkle, Physical Review {\bf{D 63}}  (2001) 044007;

\item {9.}   Lior M. Burko, Physical Review {\bf{D 62}} (2000) 127503;

\item {10.}   D.Birmingham, hep-th/0101194;

\item {11.} Gerald t'Hooft, Physical Review {\bf{D 14}} (1976) 3432;

\item {12.} A.A.Belavin, A.M.Polyakov and A.B. Zamolodchikov, Nuclear Physics,{\bf{ B241}} (1984) 333;

\item {13.} Adel Bilal, {\it{Duality in N=2 SUSY SU(2) Yang-Mills Theory}}, hep-th/9601007;

\item {14.}  N.Seiberg and E. Witten, Nuclear Physics {\bf{B426}} (1994) 19;

\item {15.} S. Carlip, Classical and Quantum Gravity, {\bf {12}} (1995) 2853;

\item {16.} Cristian Martinez, C.Teitelboim and Jorge Zanelli, Physical Review  {\bf{D61}} (2000) 1849104013.

\vfill\eject

{\bf{ FIGURE CAPTIONS}}
\vskip .2truein
Figure 1 : Plot of asymptotic behaviour of eq. 2 vs. $r$, near $r$ equals zero ($ \overline{r}$ goes to infinity).
\vskip .1truein
Figure 2 : Plot of asymptotic behaviour of eq. 3 vs. $r$, near $r$ equals zero ($ \overline{r}$ goes to infinity).
\vskip .1truein
Figure 3 : Plot of asymptotic behaviour of eq. 4 vs. $r$, near $r$ equals zero ($ \overline{r}$ goes to infinity).
\vskip .1truein
Figure 4 : Plot of asymptotic behaviour of eq. 5 vs. $r$, near $r$ equals zero ($ \overline{r}$ goes to infinity).
\vskip .1truein
Figure 5 : Plot of $-M+\overline{r}^2$ near $ \overline{r}$ equals to zero.

\vfill\eject
   \midinsert
   \vskip .75truein
   \centerline{\BoxedEPSF{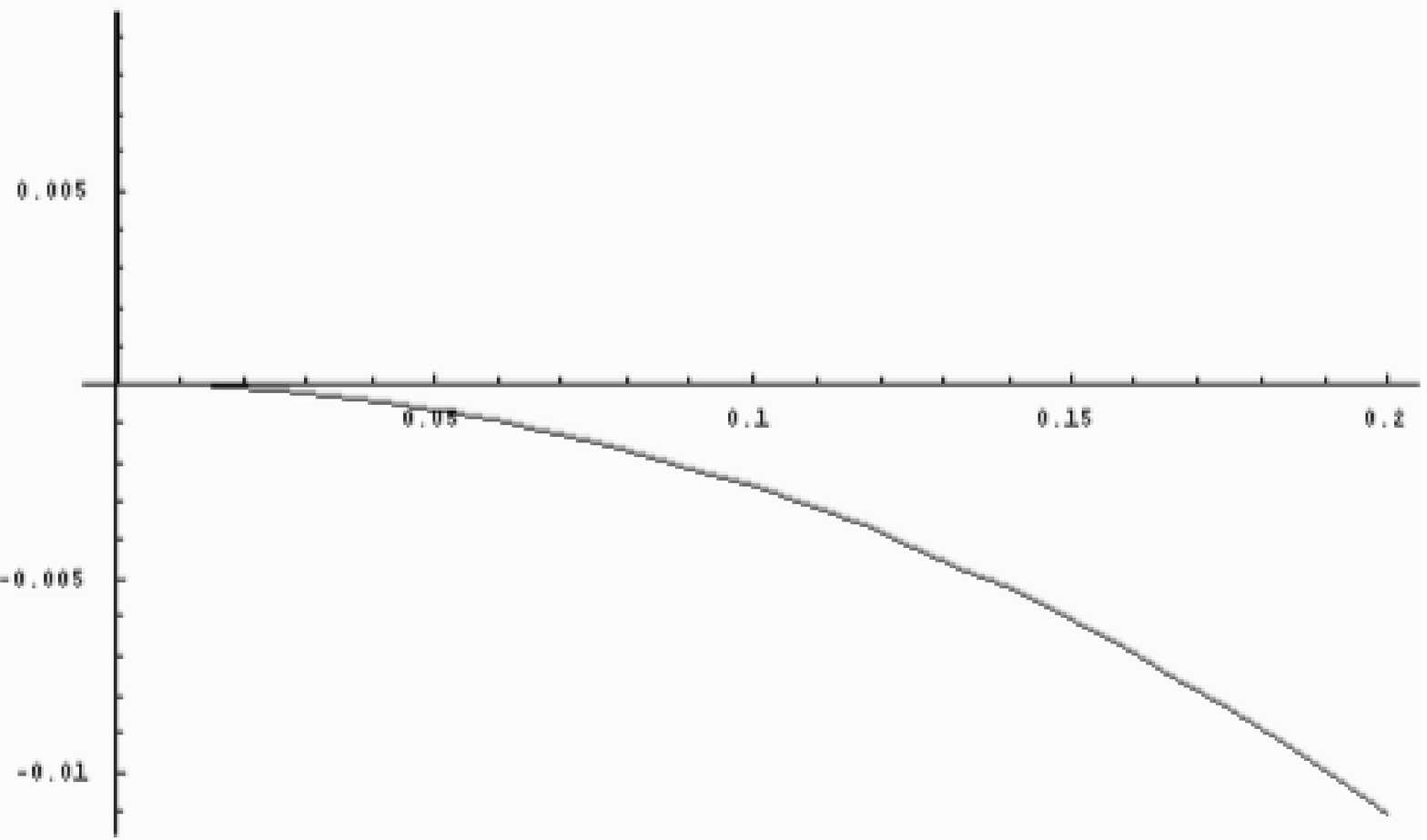}}
   \vskip .2truein
   \centerline {\bf { Figure 1}}
   \vskip .5truein
   \centerline{\BoxedEPSF{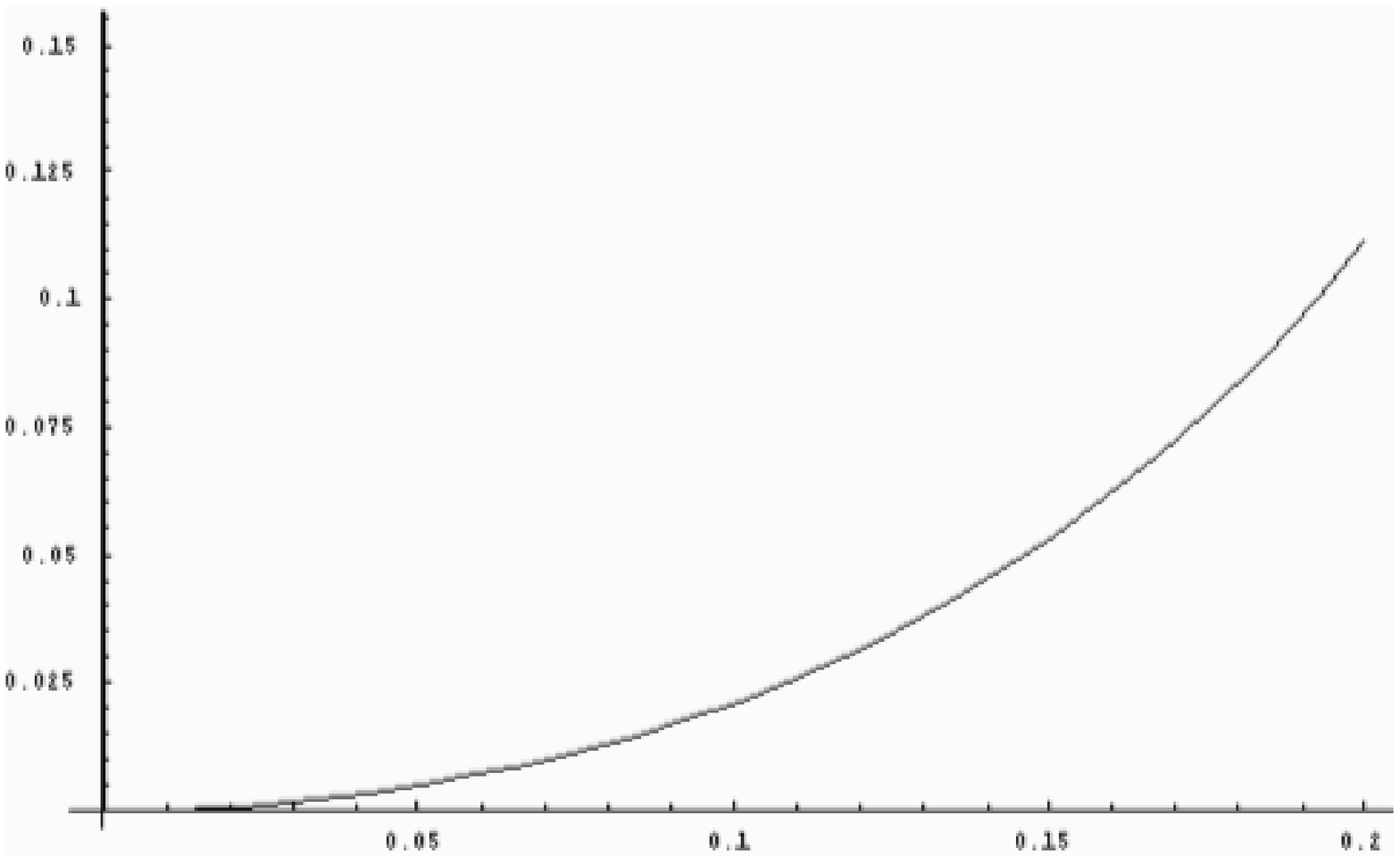}}
   \vskip .2truein
   \centerline {\bf { Figure 2}}
   \endinsert

\vfill\eject
   \midinsert
   \vskip .75truein
   \centerline{\BoxedEPSF{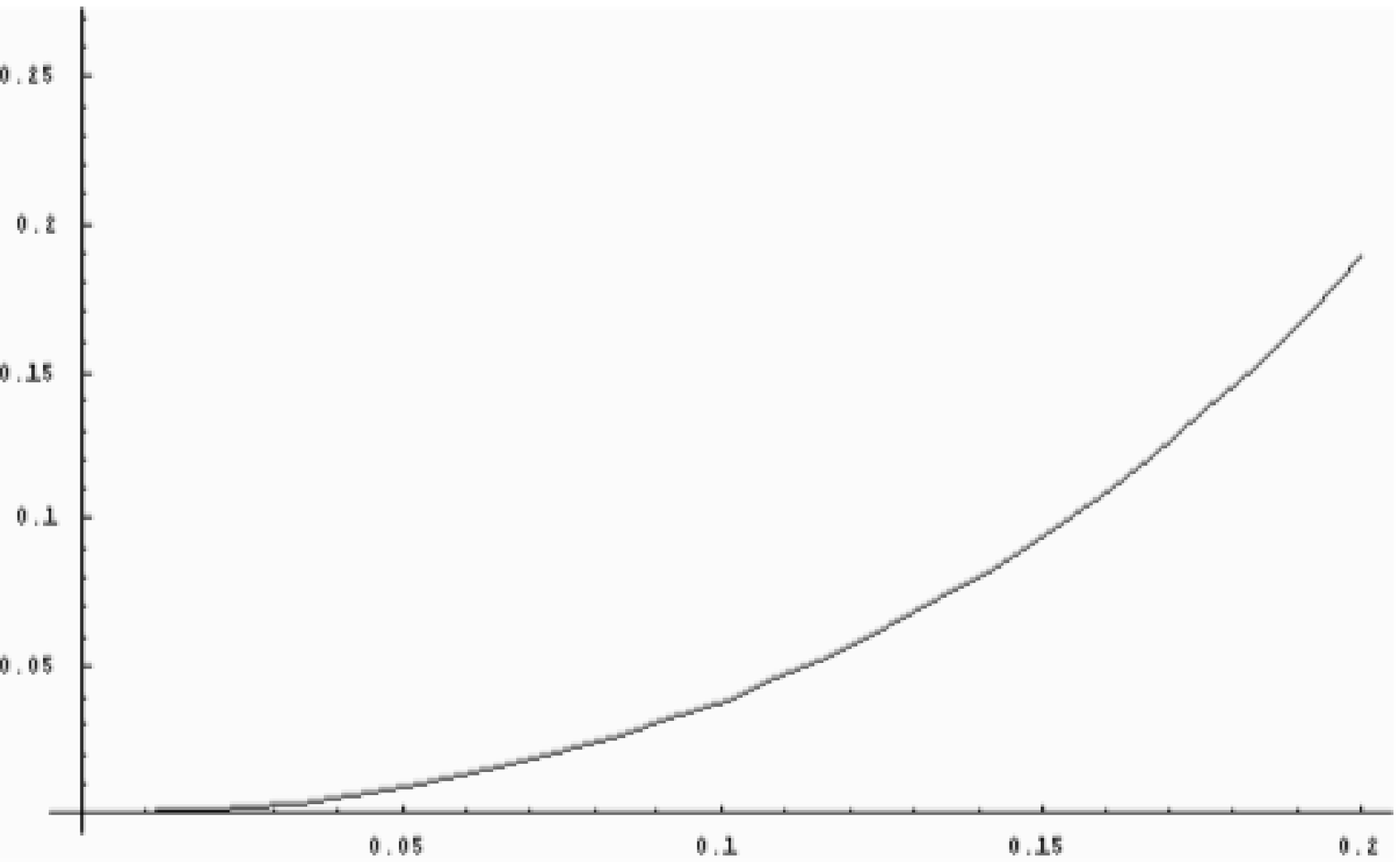}}
   \vskip .2truein
   \centerline {\bf { Figure 3}}
   \vskip .5truein
   \centerline{\BoxedEPSF{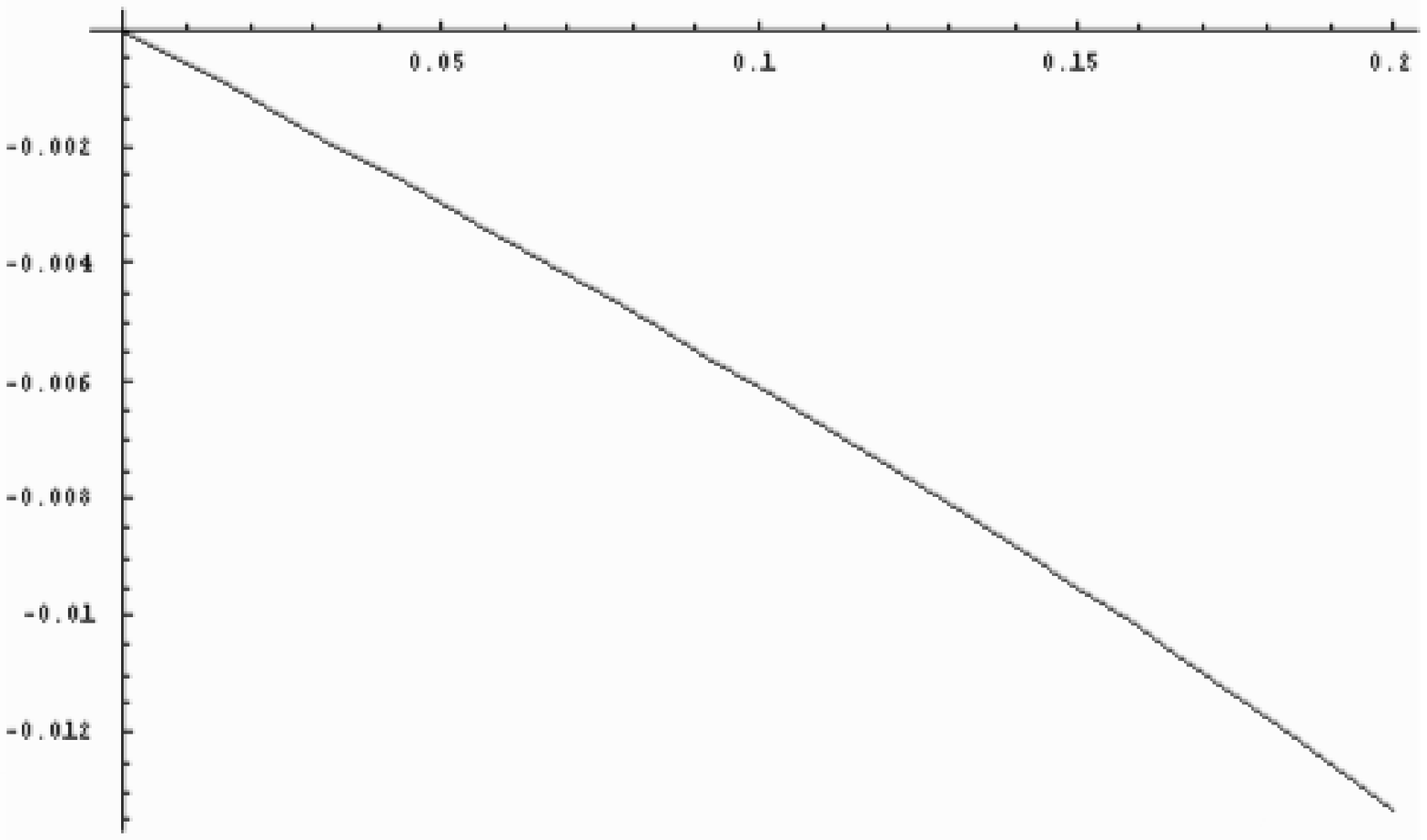}}
   \vskip .2truein
   \centerline {\bf { Figure 4}}
   \endinsert

\vfill\eject
   \midinsert
   \vskip .75truein
   \centerline{\BoxedEPSF{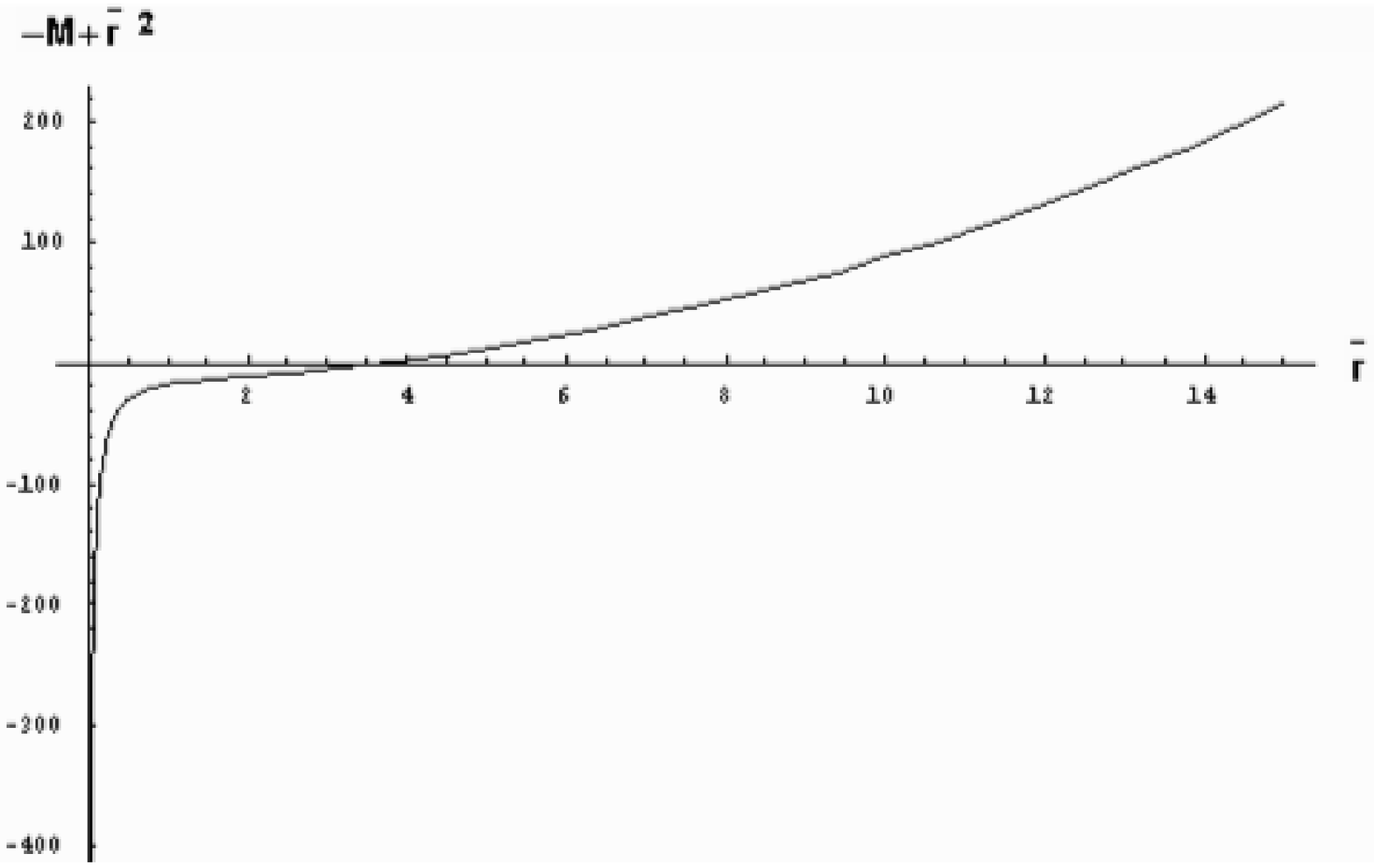}}
   \endinsert
\centerline {\bf { Figure 5}}

\end